\documentclass[12pt,a4paper]{article}

\usepackage{graphicx} 
\usepackage{amsmath}
\usepackage{amsfonts}
\usepackage{amssymb}
\usepackage{authblk}
\usepackage{a4wide}

\newcommand{\bsquare}		{$\blacksquare$}
\newcommand{\iotabar}       {\mbox{$\,\iota\!\!$-}}
\newcommand{\bxi}			{\boldsymbol{\xi}}
\newcommand {\bJ}           {\mathbf{J}}
\newcommand {\bB}           {\mathbf{B}}
\newcommand {\bE}           {\mathbf{E}}

\newcommand {\bb}           {\mathbf{b}}

\newcommand {\bC}           {\mathbf{C}}

\newcommand {\bv}           {\mathbf{v}}

\newcommand {\grad}         {\nabla}

\newcommand{\alf}			{Alfv\'{e}n }
\newcommand{\bkap}			{\boldsymbol{\kappa}}

\begin{document}
\title{Ideal MHD theory of low-frequency \alf waves in the H-1 Heliac.}

\author{J Bertram\footnote{Email contact: jason.bertram@anu.edu.au.}}
\author{B D Blackwell}
\author{M J Hole}

\affil{Plasma Theory and Modelling, Research School of Physics and Engineering, Australian National University, ACT 0200, Australia.}

\maketitle

\begin{abstract}

A part analytical, part numerical ideal MHD analysis of low-frequency \alf wave physics in the H-1 stellarator is given. The three-dimensional, compressible ideal spectrum for H-1 is presented and it is found that despite the low $\beta$ of H-1 plasmas ($\beta\approx 10^{-4}$), significant Alfv\'{e}n-acoustic interactions occur at low frequencies. Several quasi-discrete modes are found with the three-dimensional linearised ideal MHD eigenmode solver CAS3D, including beta-induced \alf eigenmode (BAE)-type modes in beta-induced gaps. The strongly shaped, low-aspect ratio magnetic geometry of H-1 causes CAS3D convergence difficulties requiring the inclusion of many Fourier harmonics for the parallel component of the fluid displacement eigenvector even for shear wave motions. The highest beta-induced gap reproduces large parts of the observed configurational frequency dependencies in the presence of hollow temperature profiles. 

\end{abstract}

\section{Introduction}

A rich low-frequency wave phenomenology showing Alfv\'{e}nic characteristics has been observed in plasmas generated on the H-1 heliac.\cite{blackwell_configurational_2009,pretty_study_2007,bertram_reduced_2011} 
Previous ideal MHD studies of these waves in cylindrical geometry has shown that cylindrical shear \alf continuum turning points reproduce important features of measured magnetic frequencies\cite{blackwell_configurational_2009,pretty_study_2007}, and cylindrical global \alf eigenmode (GAE) modelling produces plausible radial eigenmode profiles and magnetic attenuation predictions\cite{bertram_reduced_2011}. However, it is well known that \alf waves existing independently in a cylindrical plasma will couple in a toroid due to magnetic field strength modulations, and the same is true for coupling between acoustic waves and \alf waves at low-frequencies.\cite{heidbrink_basic_2008} Since these coupling effects are expected to feature prominently in low aspect-ratio, strongly-shaped plasmas like those created on H-1, this paper aims to provide a geometrically accurate treatment of H-1 wave physics within the theoretical framework of ideal MHD.

Two significant drivers of Alfv\'{e}n wave activity studies in the magnetic confinement fusion context are the direct effects such wave activity may have on plasma confinement, through energetic particle interactions for example, and the diagnostic utility wave activity provides as proxy for determining underlying plasma parameters.\cite{heidbrink_basic_2008} In particular, low-frequency Alfv\'{e}n-acoustic waves have drawn increased attention in recent years as they are generally poorly understood.\cite{kolesnichenko_sub-gam_2008,heidbrink_what_1999} Even under the simplifying assumptions of ideal MHD, the low-frequency domain is difficult to treat theoretically because, in the presence of geodesic curvature, coupling between \alf and sound branches introduces considerable complexity to the linearised eigenfrequency spectrum. Furthermore, ideal MHD provides a poor description of compressible motions\cite{freidberg_ideal_1982}, and it omits thermal ion interactions, such as ion Landau damping, which become significant in this range of frequencies.\cite{gorelenkov_beta-induced_2009} However, ideal MHD has the important advantage that it is simple enough to allow for an accurate treatment of the geometry, providing a good first model for understanding the low-frequency domain.\cite{gorelenkov_beta-induced_2009,turnbull_global_1993,eremin_beta-induced_2010}

In this work we present the three-dimensional, compressible ideal spectrum for H-1, based on numerical simulations with the three-dimensional continuous spectrum code CONTI\cite{konies_coupling_2010} in combination with the three-dimensional ideal MHD linear eigenmode solver CAS3D.\cite{schwab_ideal_1993,nuhrenberg_compressional_1999} Significant coupling of eigenmode Fourier harmonics in both toroidal and poloidal directions are induced by a substantial mirror term (inherent in a helical axis, low aspect-ratio configuration) in combination with a variety of strong field-strength modulations within a cross-section. We have found that the conditions for convergence of CAS3D in Fourier space are quite stringent for H-1 configurations, in the sense that shear \alf frequencies  dip sharply near the core unless the component of fluid motion along the field lines has a Fourier basis including at least two toroidal side-bands, even for high-frequency modes with negligible acoustic interaction. An explanation for this finding is given.

The first two helical \alf eigenmode (HAE) gaps and the toroidal \alf eigenmode (TAE) gap are identified. Significant beta-induced gap structure is found, starting at around $35~\rm{kHz}$. Consequently, H-1 fluctuations, which typically lie near or below $35~\rm{kHz}$, may be considered sub-GAM modes.\cite{kolesnichenko_sub-gam_2008} Based on analytic estimates, it is proposed that H-1 fluctuations may include beta-induced Alfv\'{e}nic eigenmodes partially reproducing characteristic ``whale-tail" structures in configuration space due to hollow temperature profiles. Low frequency discrete modes are presented.

The outline of the paper is as follows. Section 2 provides an overview low-frequency H-1 wave activity. Section 3 discusses H-1 equilibrium and magnetic geometry. Section 4 gives a summary of low-frequency stellarator \alf wave physics as well as a discussion of CAS3D convergence requirements for H-1. Section 5 presents the ideal spectrum and low-frequency discrete modes. In Section 6 we show that, in the presence of hollow temperature profiles, the highest beta-induced \alf eigenmode (BAE) gap frequency reproduces parts of the configuration-space whale-tails. We conclude in Section 7.

\section{Overview of low-frequency waves in H-1 plasmas}\label{sec:lfwavh1}

\begin{figure}[h]
\centering
\includegraphics[scale=1]{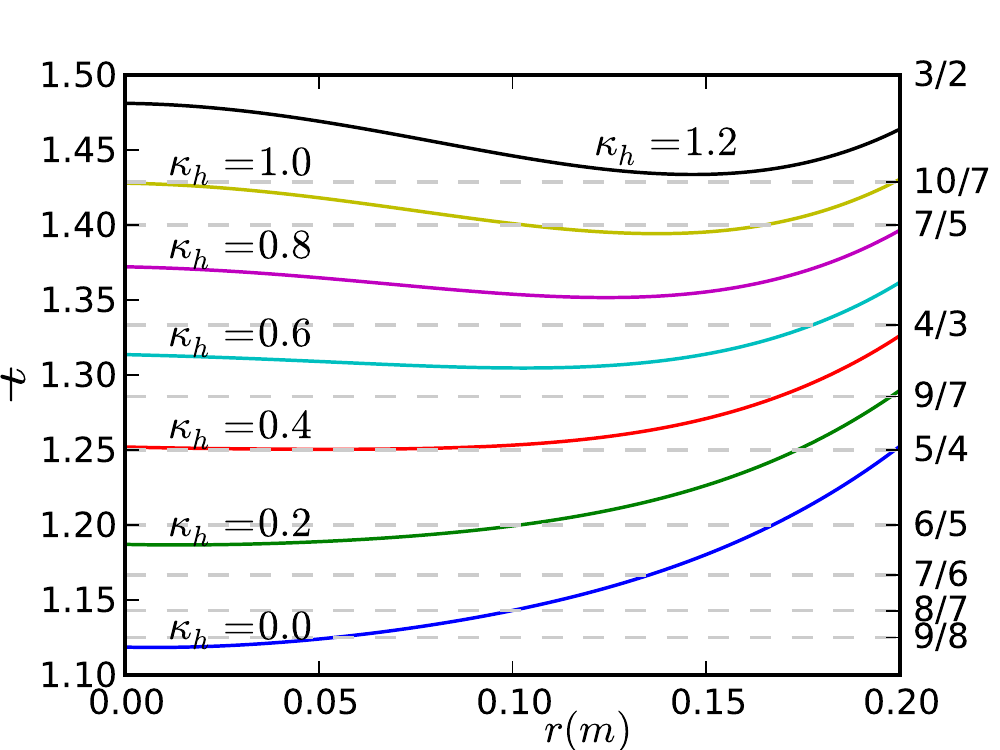}
\caption{\label{fig:iota}:  
\quad Rotational transform $\iotabar$ as a function of average minor radius for selected values of $\kappa_{\rm{h}}$, where the plasma edge lies at $r=0.2$~m. Low-order rational values are indicated on the right.}
\end{figure}

In this section we discuss low-frequency wave activity on H-1 and give a brief overview of previous studies. H-1 is a ``flexible" heliac, which means that a variety of magnetic geometries can be accessed by adjusting $\kappa_{\rm{h}}$, a parameter measuring relative coil currents (held constant during a discharge), where $0\leq\kappa_h\leq 1.2$. Different values of $\kappa_{\rm{h}}$ correspond to different plasma parameters, flux surface geometries and rotational transform $\iotabar=1/q$ profiles (see ref. \cite{blackwell_configurational_2009} for details). Perhaps the most important of these is the $\iotabar$ dependence on $\kappa_h$, shown in Figure \ref{fig:iota}.

\begin{figure}[h]
\centering
\includegraphics[scale=1]{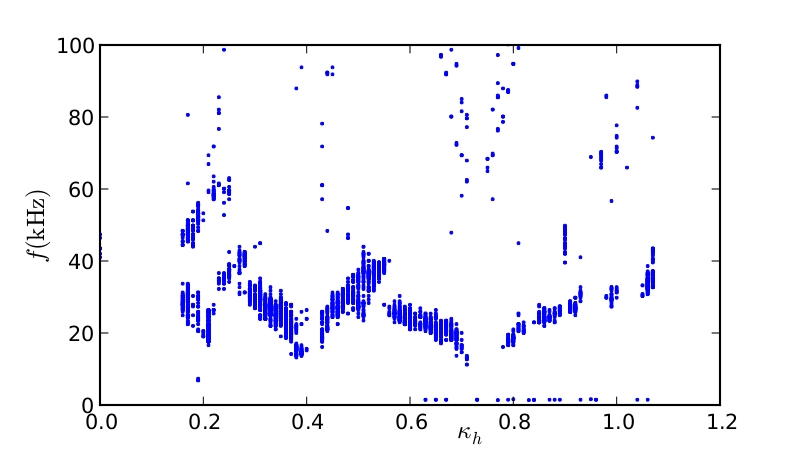}  \caption{\label{fig:freqvskappa_davethesis}:  
\quad Magnetic field fluctuation frequencies measured by poloidal Mirnov array coils against $\kappa_h$.}
\end{figure}

The configurational dependence of fluctuation frequencies is our main focus as H-1 discharges are usually dominated by a single coherent steady-frequency fluctuation present in a time interval with limited variation in equilibrium magnetic field strength and plasma density.\cite{blackwell_configurational_2009,bertram_reduced_2011} Figure \ref{fig:freqvskappa_davethesis} 
shows the frequencies measured by poloidal Mirnov coils at different times within discharges for a selection of discharges.\cite{pretty_study_2007}

It has been shown\cite{blackwell_configurational_2009,pretty_study_2007} that the ``V" structures (which we term ``whale-tails") near $\kappa_h=0.4$ and $\kappa_h=0.75$ in figure \ref{fig:freqvskappa_davethesis} \textit{qualitatively} follow the turning points of the cylindrical shear \alf frequency $\omega_{\rm{A}}=|n-\iotabar m| B_0/R \mu_0\rho$ for $(m,n)=(4,5)$ and $(m,n)=(3,4)$ respectively (compare with low-order rationals in Figure \ref{fig:iota}), where the turning points of $\omega_{\rm{A}}$ depend on $\kappa_h$ through $\iotabar$ and $\rho$ (the mass density), with the average magnetic field strength $B_0\approx 0.46~\rm{T}$ held constant during a discharge. Mode-number assignments for these two regions are obtained from poloidal Mirnov array phase signatures\cite{blackwell_configurational_2009,pretty_study_2007} and initial analysis of helical Mirnov array measurements, where $m$ and $n$ are the azimuthal and axial mode-numbers respectively. These are dominant mode-number assignments as multiple mode-numbers have not yet been resolved from the available data. Assuming the best information available for mass density and using H-1's average major radius $R=1$~m in the formula for $\omega_A$, predicted frequencies were found to be a factor of $\lambda \approx 3.7$ times larger than measured frequencies. The discrepancy factor is reduced to $\lambda \approx 3$ if the helical extension of the arclength of the magnetic axis is accounted for by setting $R=1.25$~m. This discrepancy was not the same for each of the four branches of the two whale-tails and mode localisation assumptions were made for some configurations. Consequently, the value $\lambda \approx 3$ should be taken as a coarse measure of the theory-measurement frequency discrepancy.

The frequency scaling behaviour, which is qualitatively analogous to frequency sweeping of reversed-shear \alf eigenmodes (RSAE) as $q$ changes in tokamak discharges\cite{heidbrink_basic_2008}, suggests that the low-frequency modes are GAEs and non-conventional GAEs (NGAEs) at local minima and maxima in $\omega_{\rm{A}}$ respectively. The low toroidal currents of H-1 favour the existence of NGAEs, whereas RSAEs have only been observed in tokamaks during ''Alfv\'{e}n cascades" in which several waves with different mode numbers sweep through a broad frequency range during the course of a discharge, unlike the steady, solitary H-1 modes.\cite{kolesnichenko_conventional_2007} However, no robust explanations have been found for the quantitative discrepancy. While significant impurities are likely to be present in H-1 plasmas, cylindrical frequencies are expected to depend weakly on this effect as $\omega_A$ is proportional to the inverse square root of the effective mass. Collisions with neutrals are likewise not thought to be able to explain such a large discrepancy.\cite{blackwell_configurational_2009} 

Cylindrical ideal MHD modelling of GAE eigenmodes under H-1 conditions has shown agreement with magnetic attenuation and optical emissions measurements.\cite{bertram_reduced_2011} However, as that study assumed cylindrical geometry, no comparisons were made with modes other than GAEs. Furthermore, that study highlighted the multiply-peaked nature of the Fourier spectrum of magnetic phase measurements, pointing to the importance of mode-number coupling in H-1. We will return to this ``GAE hypothesis" in view of the 3-dimensional spectrum.

\section{H-1 equilibrium with VMEC}

We use a Boozer coordinate representation $(s,\vartheta,\zeta)$ for ideal MHD equilibrium. Here $s$ denotes the normalised toroidal flux $\psi/\psi_{edge}$ where $2\pi\psi$ is the toroidal flux, and $\vartheta,\zeta$ are magnetic poloidal and toroidal Boozer angles respectively. The Variational Moments Equilibrium Code\cite{hirshman_steepest-descent_1983} (VMEC) solves the ideal equilibrium equation $\grad p = (\grad\times\bB)\times\bB$ in 3-dimensions assuming that nested flux surfaces fill the plasma volume. VMEC chooses the poloidal angle $\theta_{\rm{VMEC}}$ so that the spectral width of the Fourier representation $(R(s,\theta_{\rm{VMEC}},\phi),Z(s,\theta_{\rm{VMEC}},\phi))$ for each magnetic surface is minimised\cite{hirshman_optimized_1985}, where $(R,\phi,Z)$ are cylindrical coordinates with $\phi$ the geometric toroidal angle. This yields significant computational benefits, although further computation is required to obtain the larger magnetic coordinate Fourier representation $(R(s,\vartheta,\zeta),Z(s,\vartheta,\zeta))$.

We have opted to give VMEC predetermined $\iotabar$ profiles since the profiles shown in Figure \ref{fig:iota}, which are vacuum field profiles, are considered reliable when a plasma is present due to the fact that H-1 operates at low-$\beta$ ($\beta\approx 10^{-4}$).\cite{glass_tomographic_2004} Pressure profiles are obtained from electron density measurements using the ideal gas law $p=n_e k_B T_e+n_i k_B T_i=2 n k_B T_e$ assuming $n_i=n_e$ and constant plasma temperatures $T_i=T_e=20$~eV.\cite{pretty_study_2007} We have run VMEC in fixed-boundary mode, with boundary surfaces obtained using DESCUR which generates the  representation $(R(s,\theta_{\rm{VMEC}},\phi),Z(s,\theta_{\rm{VMEC}},\phi))$ from a field-line trace on a magnetic surface.\cite{hirshman_optimized_1985} The field-line trace has field periodicity of $N=3$ (i.e. all equilibrium quantities are 3-periodic in the toroidal direction) and is stellarator symmetric
\begin{equation}
R(s,\vartheta,\zeta)=R(s,-\vartheta,-\zeta) \quad \quad Z(s,\vartheta,\zeta)=-Z(s,-\vartheta,-\zeta)
\end{equation} 
as symmetry-breaking terms arising from the response of the magnetic field coils to the applied field are ignored. 

\begin{figure}[h]
\centering
$\begin{array}{cc}
(a)&(b)\\
\includegraphics[scale=1]{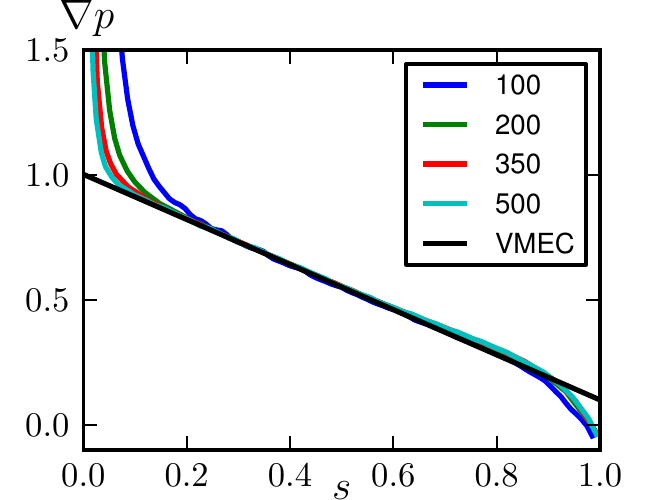} & 
\includegraphics[scale=1]{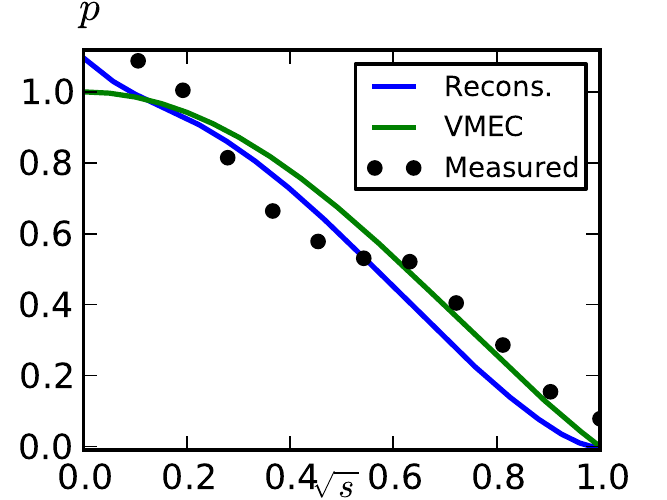}  
\end{array}$
\caption{\label{fig:mc3d_press}:  
\quad  (a) Prescribed VMEC pressure gradient compared with the MC3D-reconstructed pressure gradient for increasing numbers of flux surfaces $N_s=100,200,350,500$. (b) Pressure reconstructed from the $N_s=500$ case in (a), compared with the prescribed VMEC pressure as fitted to electron density inversions\cite{oliver_electronically_2008}. All pressures normalised to core fitted pressure. This is a $\kappa_h=0.54$ configuration.}
\end{figure}

Transformation to Boozer coordinates is performed with the MC3D code (part of the CAS3D package). This transformation gives poorly reconstructed pressure profiles (from $\grad p = (\grad\times\bB)\times\bB$ where  $\bB$ is reconstructed in the Boozer coordinate Fourier basis) for H-1 configurations unless a VMEC equilibrium with a large number magnetic surfaces $N_s$ is supplied. Figure \ref{fig:mc3d_press}a shows the improvement of the agreement between the prescribed VMEC pressure gradient and the MC3D-reconstructed pressure gradient with increasing $N_s$. The reconstructed pressure for the $N_s=500$ case is shown in figure \ref{fig:mc3d_press}b along with the prescribed VMEC pressure and electron density inversions from a $\kappa_h=0.54$ discharge. Singular behaviour at the core is a consequence of the $s=0$ coordinate singularity arising from the vanishing   of the poloidal flux gradient, which affects the VMEC equilibrium as well as the Boozer coordinate transformation.\cite{hirshman_steepest-descent_1983} The $\grad p$ discrepancy near the edge may be due to the difficulty of representing $\bB$ on the more strongly shaped outer magnetic surfaces, even though the geometry of the flux surfaces is accurately reconstructed. The largest $N_s=500$ Fourier table used before memory requirements became prohibitive was $0\leq m \leq 40$ and $-30\leq n \leq 30$. Figure \ref{fig:mc3d_press}a shows little improvement moving from $N_s=350$ to $N_s=500$, although agreement is harder to obtain for higher $\beta$ values (for example $\beta=5\times 10^{-4}$ compared with the present $\beta=1\times 10^{-4}$) where $N_s=500$ flux surfaces or more are necessary (larger $\beta$ values are expected with the recently upgraded H-1 heating system\cite{blackwell_australian_2010}).

\begin{figure}[h]
\centering
$$\setlength\arraycolsep{0.0em}
\begin{array}{cc}
(a)&(b)\\
\includegraphics[scale=1]{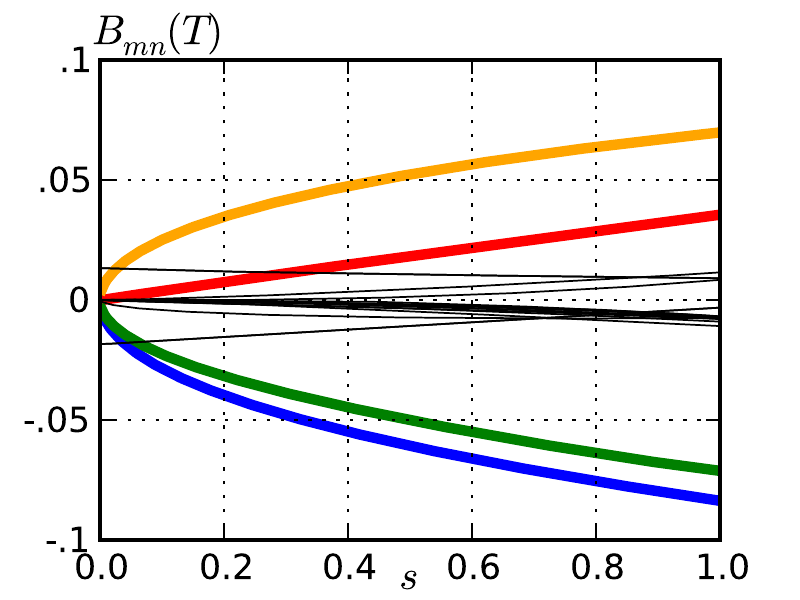} &
\includegraphics[scale=1]{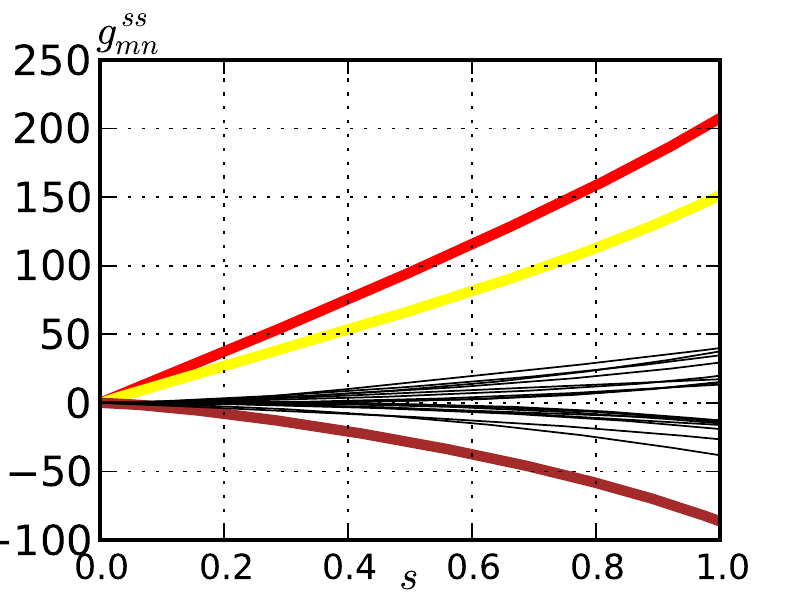} 
\end{array}$$
\caption{\label{fig:bmngss_mc}:  
\quad (a) Fourier components of the magnetic field strength $B=|\bB|$ in Boozer coordinates on a field period. The largest four components shown with thick lines are: $(m,n)=(0,0)$-red (with $B_0=0.46$ subtracted), $(1,0)$-blue, $(1,-1)$-green, $(1,-2)$-orange. All other components in thin black. (b) Fourier components of the contravariant metric component $g^{ss}\equiv \grad s \cdot \grad s$ in magnetic coordinates. The largest three components shown with thick lines are: $(m,n)=(0,0)$-red, $(2,-2)$-yellow, $(3,-3)$-brown. All other components in thin black. This is a $\kappa_h=0.54$ configuration.}
\end{figure}

Stellarator symmetry simplifies the Fourier representation of equilibrium quantities by permitting sine-only or cosine-only expansions. The Boozer coordinate Fourier expansion of $B=|\bB|$ on a field period can be written as\cite{hirshman_steepest-descent_1983,kolesnichenko_conventional_2007}
\begin{equation}
B(s,\vartheta,\zeta)= \sum_{mn} B_{mn}(s) \cos(m\vartheta+nN\zeta)=B_0 \left(1+\sum_{mn} b_{mn}(s) \cos(m\vartheta+nN\zeta)\right). \label{eq:Bfour}
\end{equation} 
Figure \ref{fig:bmngss_mc}a gives the flux surface dependence of the $B_{mn}$ for $\kappa_h=0.54$ in magnetic coordinates showing that $m=1$ Fourier components dominate the poloidal field strength modulation. Similarly, the Boozer coordinate Fourier representation of the contravariant metric component $g^{ss}\equiv \grad s \cdot \grad s$ can be written as
\begin{equation}
g^{ss}(s,\vartheta,\zeta)= \sum_{mn} g^{ss}_{mn}(s) \cos(m\vartheta+nN\zeta)=\delta_0 \frac{2 \psi B_0}{\psi_{edge}^2} \left(1+\sum_{mn} \tilde{g}^{ss}_{mn}(s) \cos(m\vartheta+nN\zeta)\right). \label{eq:gfour}
\end{equation} 
where the last expression on the right defines $\tilde{g}^{ss}_{mn}(s)$ and $\delta_0=\delta_0(\psi)$ which is related to the elongation of flux surface cross-sections.\cite{kolesnichenko_conventional_2007} The flux surface dependence of the $g^{ss}_{mn}$ are shown in Figure \ref{fig:bmngss_mc}b, illustrating the strong ellipticity ($m=2$) and triangularity ($m=3$) of the bean-shaped cross-sections. 

\section{Wave Theory Preliminaries}

\subsection{Mode Families} \label{sec:modefam}

The toroidal $N$-periodicity of stellarators leads to the notion of independent mode-families. A choice of mode family is the stellarator equivalent to a choice of toroidal mode number $n$ for tokamaks, as different $n$ do not couple in axisymmetric geometry. 

The condition for coupling between Fourier mode-numbers can be obtained from a variational formulation of the linearised ideal MHD equations.\cite{bernstein_energy_1958} Eigenvalues  $\omega$ and plasma displacement eigenfunctions $\bxi$ of the linearised ideal MHD equations extremize the Lagrangian 
\begin{equation}
\mathcal{L}=\omega^2 K - \delta W
\end{equation}
with respect to variations of $\bxi$\cite{bernstein_energy_1958}, where the kinetic energy is $K=\frac{1}{2}\int\rho|\bxi|^2d\mathbf{r}$ and the potential energy $\delta W$ can be written as\cite{schwab_ideal_1993}
\begin{equation}
\delta W=\frac{1}{2}\int |\bC|^2-D(\bxi\cdot\grad s)^2+\gamma p (\grad\cdot\bxi)^2 d\mathbf{r}, \label{eq:Wpotcas3d}
\end{equation}
where
\begin{equation}
\bC=\grad\times(\bxi\times\bB)+\frac{\bJ\times\grad s}{|\grad s|^2}\bxi\cdot\grad s,
\end{equation}
\begin{equation}
D=2\frac{(\bJ\times\grad s)\cdot(\bB\cdot\grad)\grad s}{|\grad s|^4},
\end{equation}
and boundary and vacuum energy terms have been neglected i.e. the plasma is assumed to be surrounded by a perfectly conducting wall. The latter assumption is made for simplicity, as only fixed boundary equilibria are considered here, and it does not affect the mode family argument that follows.

All terms in the $\delta W$ integrand are of the form  ``perturbation $\times$ perturbation $\times$ equilibrium". Substituting the general (combined cosine and sine) Fourier expansion for $\bxi$ (with poloidal mode-numbers $m$ and toroidal mode-numbers $n$) and the general Fourier expansion for each equilibrium quantity (where the toroidal mode-numbers have the form $nN$), the orthogonality of the trigonometric functions leads to the condition for coupling
\begin{equation}
n+n'\in N\mathbb{Z}  \quad \mbox{or} \quad  n-n'\in N\mathbb{Z}, \label{eq:couplingcond}
\end{equation}
where $N\mathbb{Z}=\{\ldots,-2N,-N,0,N,2N,\ldots\}$.\cite{schwab_ideal_1993}

The $N=3$ toroidal periodicity of H-1 configurations implies the existence of two independent mode-families $N=0,1$, shown in Table \ref{tab:modefam}, which are uncoupled. We will focus on the $N=1$ mode family since it contains both of the dominant mode-number pairs $(4,5)$ and $(3,4)$ for the two whale-tails (see section \ref{sec:lfwavh1}).
 
\begin{table}
\centering
\begin{tabular}{| c | c | c | c | c | c | c | c | c | c | c | c | c | c | c | c |}
\hline
$n$ & $\ldots$ & -6 & -5 & -4 & -3 & -2 & -1 & 0 & 1 & 2 & 3 & 4 & 5 & 6 & $\ldots$ \\ \hline
$N$ & $\ldots$ & \bsquare & $\square$ & $\square$ & \bsquare & $\square$ & $\square$ & \bsquare & $\square$ & $\square$ & \bsquare & $\square$ & $\square$ & \bsquare & $\ldots$ \\
\hline
\end{tabular}
\caption{\label{tab:modefam}:  
\quad Summary of H-1 mode-families with $N=0$ represented by $\blacksquare$ and $N=1$ by $\square$.} 
\end{table}

\subsection{Continuous Spectrum}\label{sec:conttheory}

The continuous ideal MHD spectrum for a general three-dimensional toroid can be obtained by CONTI\cite{konies_coupling_2010} which solves the coupled continuum equations\cite{cheng_low-n_1986}
\begin{align}
\left[\frac{\omega^2\rho|\grad\psi|^2}{B^2}+(\bB\cdot\grad)\left(\frac{|\grad\psi|^2}{B^2}(\bB\cdot\grad)\right)\right]\xi_s+\gamma p\kappa_s(\grad\cdot\bxi)=0,\label{eq:alfcont}\\
\kappa_s\xi_s+\left[\frac{\gamma p + B^2}{B^2}+\frac{\gamma p}{\omega^2\rho}(\bB\cdot\grad)\frac{1}{B^2}(\bB\cdot\grad)\right](\grad\cdot\bxi)=0, \label{eq:soundcont}
\end{align}
where $\gamma$ is the adiabatic index, $\omega$ is the angular frequency, $\xi_s=\bxi\cdot(\bB\times\grad\psi)/|\grad\psi|^2$ is the surface displacement and  $\kappa_s=2\bkap\cdot(\bB\times\grad\psi)/B^2$ is the geodesic curvature, which is proportional to the surface component $(\grad\psi\times\bb)\cdot\bkap/B$ of the magnetic field line curvature $\bkap \equiv (\bb\cdot\grad\bb)$ where $\bb=\bB/B$. The continuum equations are derived from the linearised ideal MHD equations. For a given flux surface $\psi$ they give the condition for the existence of non-square-integrable eigenfunctions with a spatial singularity at $\psi$ (see ref. \cite{cheng_low-n_1986} for details and further references). 

If $\kappa_s$ vanishes, equations (\ref{eq:alfcont}) and (\ref{eq:soundcont}) decouple and describe the shear \alf and compressional (acoustic and compressional Alfv\'{e}n) branches of the ideal MHD continuum respectively.  Coupling between Fourier components of the shear \alf spectrum arises due to modulation in $|\grad\psi|/B$, where Fig. \ref{fig:bmngss_mc} shows the numerator (up to a constant factor) and denominator in this expression.

In the presence of nonzero $p$ and $\kappa_s$, equations (\ref{eq:alfcont}) and (\ref{eq:soundcont}) couple. Setting aside the high-frequency compressional \alf waves, it can be seen from the identity 
\begin{equation}
\grad\cdot\bxi=\bb\cdot\grad \xi_\parallel + \grad\cdot\bxi_\perp \label{eq:divxidecom}
\end{equation}
that this coupling involves two factors.\cite{breizman_plasma_2005} The first term on the right represents coupling between the acoustic motion parallel to the magnetic field lines and the surface displacement $\xi_s$. The second term represents the compressional response due to motion across the equilibrium field lines $\bv_\perp=\bB\times\bE/B^2$ in the presence of geodesic curvature.\cite{turnbull_global_1993} This effect is closely related to the electrostatic geodesic acoustic mode (GAM).\cite{winsor_geodesic_1968} 

Due to the mixing of shear and compressional motion at low frequencies the clean division between them breaks down and it is necessary to introduce classification schemes. Three schemes are used in this paper, classifying eigenfunctions as shear Alfv\'{e}nic if: (1) the largest Fourier harmonic of $\grad\cdot\bxi$ is larger than the largest Fourier harmonic for $\xi_s$ in Boozer coordinates (CONTI classification), (2) the parallel component $K_\parallel=\frac{1}{2}\int\rho|\xi_{\parallel}|^2d{\mathbf{r}}$ of the displacement kinetic energy $K$ dominates i.e. $K_\parallel/K > 0.5$ (polarisation classification) and (3) the plasma compression energy $\frac{1}{2}\int\gamma p(\grad\cdot\bxi)^2d{\mathbf{r}}$ dominates the other terms (field line bending and compression energies) in $\delta W$ (potential energy classification). 

A criterion for estimating the strength of the Alfv\'{e}n-acoustic interaction has been proposed, given in terms of a ``sound parameter", \cite{kolesnichenko_conventional_2007}
\begin{equation}
\mathfrak{G}=\frac{\iotabar^2}{2}\frac{\delta_0\epsilon^2}{b^2_{10}},
\end{equation}
where $\epsilon=\sqrt{2\psi/B_0}/R$ is a measure of the inverse aspect ratio (since the toroidal flux is $2\pi\psi=B_0\pi r^2$ in a torus) and  $b_{10}$ and $\delta_0\geq 1$ are defined in equations (\ref{eq:Bfour}) and (\ref{eq:gfour}) respectively. If $\mathfrak{G}\gg 1$, the interaction is predicted to be weak and the \alf frequency is zero at the surface where $n-\iotabar m = 0$ as expected for the incompressible shear \alf spectrum. On the other hand, if $\mathfrak{G}\ll 1$, significant deviations from the incompressible shear \alf spectrum are expected with the formation of Alfv\'{e}n-sound gaps and in some cases, immediately above these, $\beta$-induced gaps bounded above by geodesic acoustic frequencies $\omega_G$ located near the surface where $n-\iotabar m = 0$.\cite{kolesnichenko_conventional_2007} For H-1 the sound parameter ranges from $\mathfrak{G}\approx 0.75$ at $\kappa_h=0.3$ to $\mathfrak{G}\approx 1.1$ at $\kappa_h=0.9$ suggesting non-trivial Alfv\'{e}n-acoustic interactions at low frequencies. H-1 is unusual in this regard, since $\epsilon/b_{10}>1$ for most stellarators (H-1 has $\epsilon/b_{10}<1$), and therefore it is often the case that $\mathfrak{G} \gg 1$ for stellarators with $\iotabar>1$.\cite{kolesnichenko_conventional_2007}

An approximate expression for the Alfv\'{e}n resonance ignoring Alfv\'{e}n-\alf interactions (equation (48) in ref. \cite{kolesnichenko_conventional_2007}) can be used to obtain estimates for $\omega_G$ by taking the limit $n-\iotabar m \rightarrow 0$ yielding for H-1
\begin{align}
\frac{b_{1,0}^2}{\omega^2-\iotabar^2 c_s^2/R^2}+\frac{b_{1,-1}^2}{\omega^2-(N-\iotabar)^2 c_s^2/R^2}+  \frac{b_{1,-2}^2}{\omega^2-(2N-\iotabar)^2 c_s^2/R^2}=\frac{2\psi\delta_0}{B_0 c_s^2} \label{eq:gamfreqest},
\end{align}
where all flux-surface quantities are evaluated at the surface where $n-\iotabar m = 0$, $N=3$, $c_s=\sqrt{\gamma p/\rho}$ is the adiabatic sound speed, $\delta_0 \sim 1.5$ and only the largest harmonics identified in Fig. \ref{fig:bmngss_mc} with thick lines are considered. The geodesic frequencies $\omega_G$ form a subset of the solutions $\omega$ to equation (\ref{eq:gamfreqest}).\cite{kolesnichenko_conventional_2007} 

\subsection{Discrete Spectrum and CAS3D convergence}

The full ideal MHD linearised eigenmode spectrum, including discrete or quasi-discrete modes existing within gaps in the continuous spectrum, can be found with CAS3D\cite{schwab_ideal_1993,nuhrenberg_compressional_1999} which solves the eigenvalue problem $\delta\mathcal{L}=0$ (see section \ref{sec:modefam}).

For H-1 configurations, CAS3D produces poorly converged spectra with shear \alf frequencies dipping sharply near the core unless the parallel displacement $\xi_\parallel$ is represented by many Fourier harmonics. Some insight into this problem can be obtained by rearranging equation (\ref{eq:divxidecom}) and solving in magnetic coordinates to give
\begin{equation}
\xi_{\parallel mn} = \frac{-i [B\sqrt{g}(\grad\cdot\bxi-\grad\cdot\bxi_{\perp})]_{mn}}{n-\iotabar m},
\end{equation}
where the Jacobian is $\sqrt{g}=[\grad(\psi)\cdot(\grad\vartheta\times\grad\zeta)]^{-1}$. For a physical divergence $\grad\cdot\bxi$ that needs to be numerically approximated, it can be seen that the spectral width of $\xi_{\parallel}$ must be larger than that of the other displacement components due to the modulation in $B\sqrt{g}$. CAS3D uses Boozer coordinates for which $B\sqrt{g}\propto 1/B$.

Inadequate representation of $\xi_\parallel$ has deleterious effects well outside of the frequency domain in which parallel ion-acoustic motion is physically important. For simplicity, let us consider the case of approximately ``pure" shear \alf waves which should have $\grad\cdot\bxi\approx 0$ and $\xi_\parallel\approx0$. The parallel displacement $\xi_\parallel$ appears only in the fluid compression term $\gamma p |\grad\cdot\bxi|^2$ and the parallel component of the kinetic energy $K_\parallel=\frac{1}{2}\int\rho|\xi_{\parallel}|^2d{\mathbf{r}}$. If $\xi_\parallel$ is not represented with a sufficient number of Fourier harmonics, both $\grad\cdot\bxi\approx 0$ and $\xi_\parallel\approx0$ will not be well satisfied, and both $\int\gamma p(\grad\cdot\bxi)^2d{\mathbf{r}}$ and $K_{\parallel}$ will be larger than they should be. Although $\gamma p |\grad\cdot\bxi|^2$ will still be small relative to the magnetic terms in $\delta W$ when $\beta$ is small, an inflated $K_{\parallel}$ will cause significant distortions to the shear \alf spectrum. 

This effect has been documented for unstable modes with the 3-dimensional ideal spectral code TERPSICHORE.\cite{fu_fully_1992} For CAS3D with W7-X equilibria it was found that a one-third larger Fourier table for the parallel displacement was sufficient to avoid these problems.\cite{nuhrenberg_compressional_1999} For H-1, due to a significant mirror-term modulation of around $20\%$, CAS3D \alf spectra are grossly distorted near the core where poloidal couplings are small and $\grad\cdot\bxi\approx 0$ cannot locally be satisfied with the inadequate ``toroidal representation" of $\xi_\parallel$, unless the first and second toroidal side-bands $n-2N$, $n-N$, $n+N$ and $n+2N$ of near-resonance harmonics $n$ are included for $\xi_\parallel$. The inclusions of these side-bands and their poloidal neighbours leads to Fourier tables twice or three times as large as those for the other components. Our simulations with H-1-like reduced-mirror-term configurations which retain strong poloidal modulations has shown that only the first toroidal side-band is required for $\xi_\parallel$.

\section{Numerical Results}

This section starts with a description of the broad features of the H-1 linearised spectrum before moving on to a more detailed study of the low-frequency part of the spectrum. Configuration space can be divided into four regions of interest corresponding to the left and right sides of the two whale-tails. On the left side of the low-$\kappa_h$ whale-tail, the $(m,n)=(4,5)$ shear \alf continuum intersects the low-frequency domain because there is a flux surface in the plasma with $\iotabar=5/4$ whereas on the right side of the low-$\kappa_h$ whale-tail there is no surface for which $\iotabar=5/4$ and the $(m,n)=(4,5)$ shear \alf continuum rapidly moves to high frequencies with increasing $\kappa_h$. The same pattern repeats itself for the $(m,n)=(3,4)$ shear \alf continuum and the  high-$\kappa_h$ whale-tail. Consequently, we focus our attention on $\kappa_h=0.30$ as representative of low-frequency Alfv\'{e}n-acoustic interactions on H-1. 

\subsection{\alf spectrum}

\begin{figure}[h]
\centering
$$\setlength\arraycolsep{0 em}
\begin{array}{cc}
(a)&(b)\\
\includegraphics[scale=1]{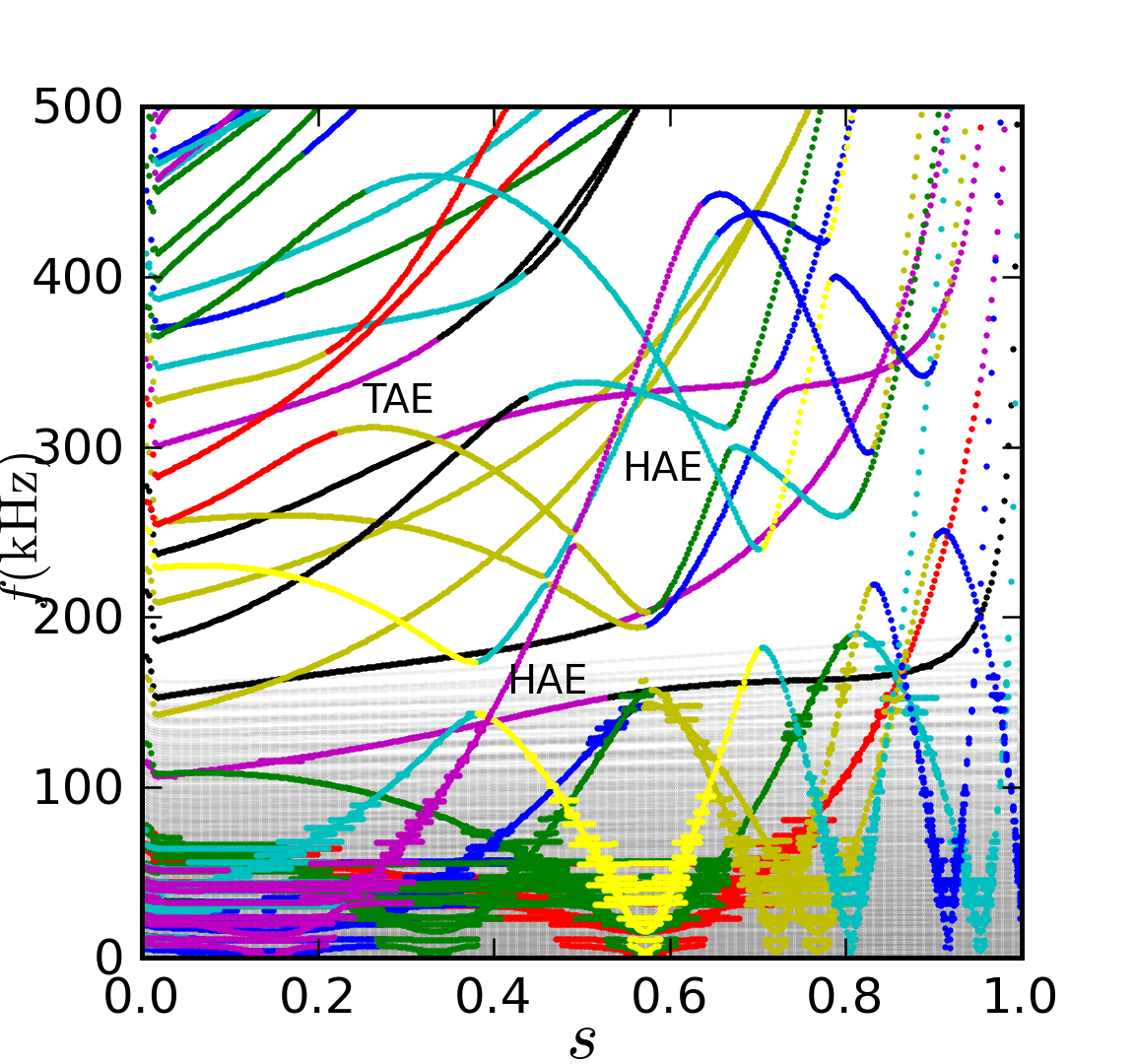}   &
\includegraphics[scale=1]{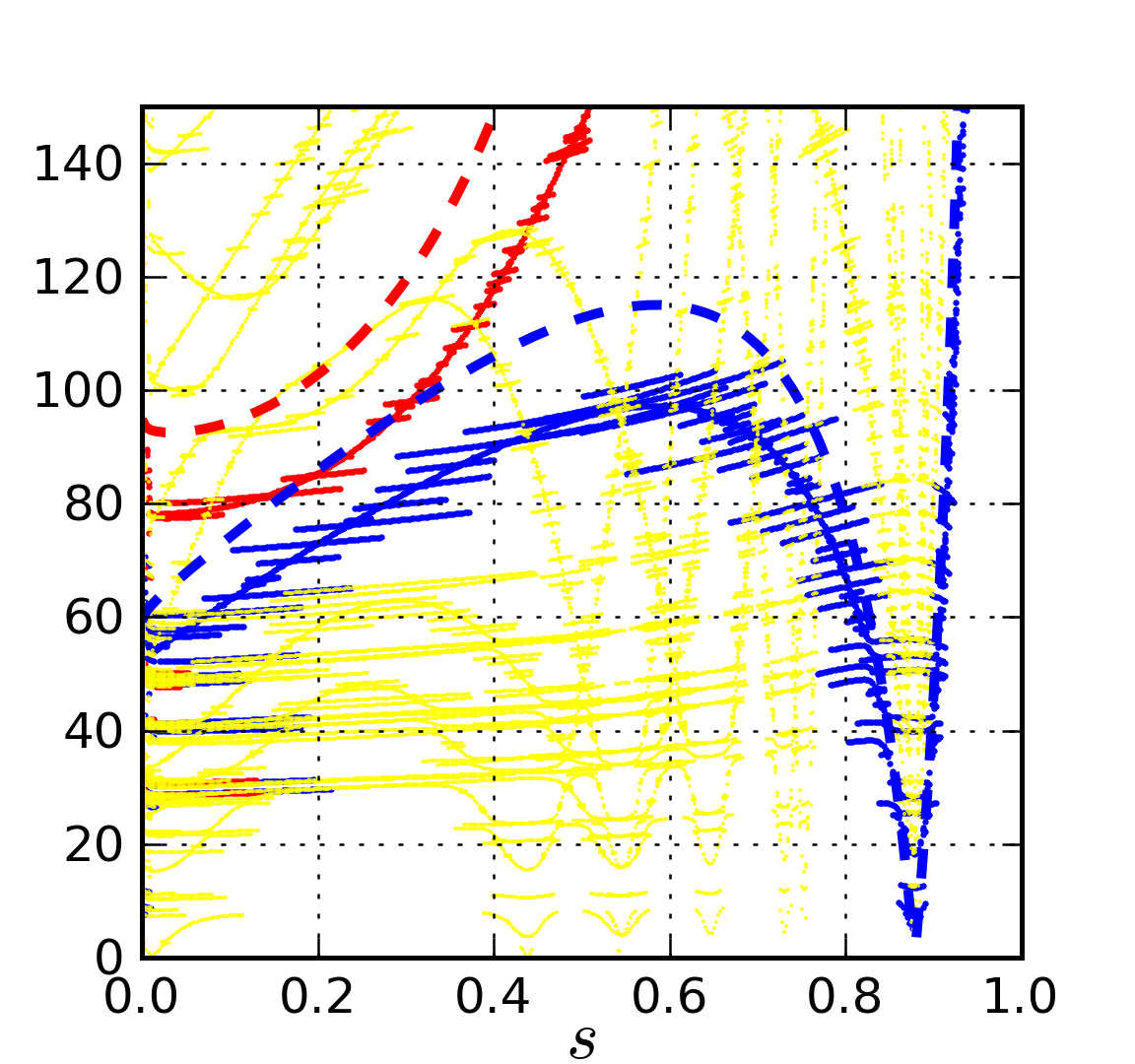}  
\end{array}$$
  \caption{\label{fig:contispect_30_full}:  
\quad (a) CONTI continuous spectrum for $N=1$, $\kappa_h=0.30$ with with $p$ and $\rho$ linear in $s$ and constant temperature. Sound branches in grey, \alf branches in colour. (b) Continuous spectrum for $N=1$, $\kappa_h=0.54$ with $p$ and $\rho$ quadratic in $s$ as in Fig. \ref{fig:mc3d_press}b and constant temperature. Modes classified as acoustic not shown, \alf branches in yellow apart from $(4,5)$-red and $(3,4)$-blue with dashed lines showing corresponding cylindrical frequencies. For both figures physical parameters are $B=0.46$, core mass density $\rho_0=1\times 10^{-8}~\rm{kg/m^3}$, core pressure $p_0=16~\rm{Pa}$ and CONTI numerical parameters are $400$ flux surfaces, $1\leq m\leq 18$ and $-2 \leq n \leq 23$ (with multiples of $3$ removed to exclude N=0).}
\end{figure}

The shear \alf gap structure for a $\kappa_h=0.30$ H-1 configuration obtained with CONTI can be seen in Figure \ref{fig:contispect_30_full}a. Two HAE gaps are present, a large one at around $150$~kHz corresponding to a coupling displacement of $(\Delta m,\Delta n)=(2,3)$ (i.e. coupling between $(m,n)$ and $(m+\Delta m,n+\Delta n)$), and above it, a small gap corresponding to $(\Delta m,\Delta n)=(3,3)$. Above both of these gaps lies the TAE gap at around $350~\rm{kHz}$. This overall gap structure is found to be qualitatively the same for all $\kappa_h$, with significant quantitative differences when there are large changes in $\rho$ near $\kappa_h=0.4$ and $\kappa_h=0.75$.\cite{blackwell_configurational_2009} These gaps are far too high to offer an explanation for the observed low-frequency wave activity.

Figure \ref{fig:contispect_30_full}b shows a comparison of the $(4,5)$ and $(3,4)$ \alf branches with their cylindrical equivalents for $\kappa_h=0.54$ (at the overlap of the two whale-tails). This shows that apart from the appearance of gap structure, at frequencies greater than $50~\rm{kHz}$ the 3-dimensional shear \alf frequency turning points are about a factor of $0.85$ lower than corresponding cylindrical frequency turning points, implying a reduction of the GAE-measurement frequency discrepancy to $\lambda \approx 2.5$ (see section \ref{sec:lfwavh1}). Geometric effects therefore alter GAE frequencies more than previously thought\cite{blackwell_configurational_2009,pretty_study_2007}, but are not sufficient to bring them in line with measurements.

\subsection{Alfv\'{e}n-acoustic spectrum}\label{alf_acoustic}

\begin{figure}[h]
\centering
$$\setlength\arraycolsep{0 em}
\begin{array}{cc}
(a)&(b)\\
\includegraphics[scale=1]{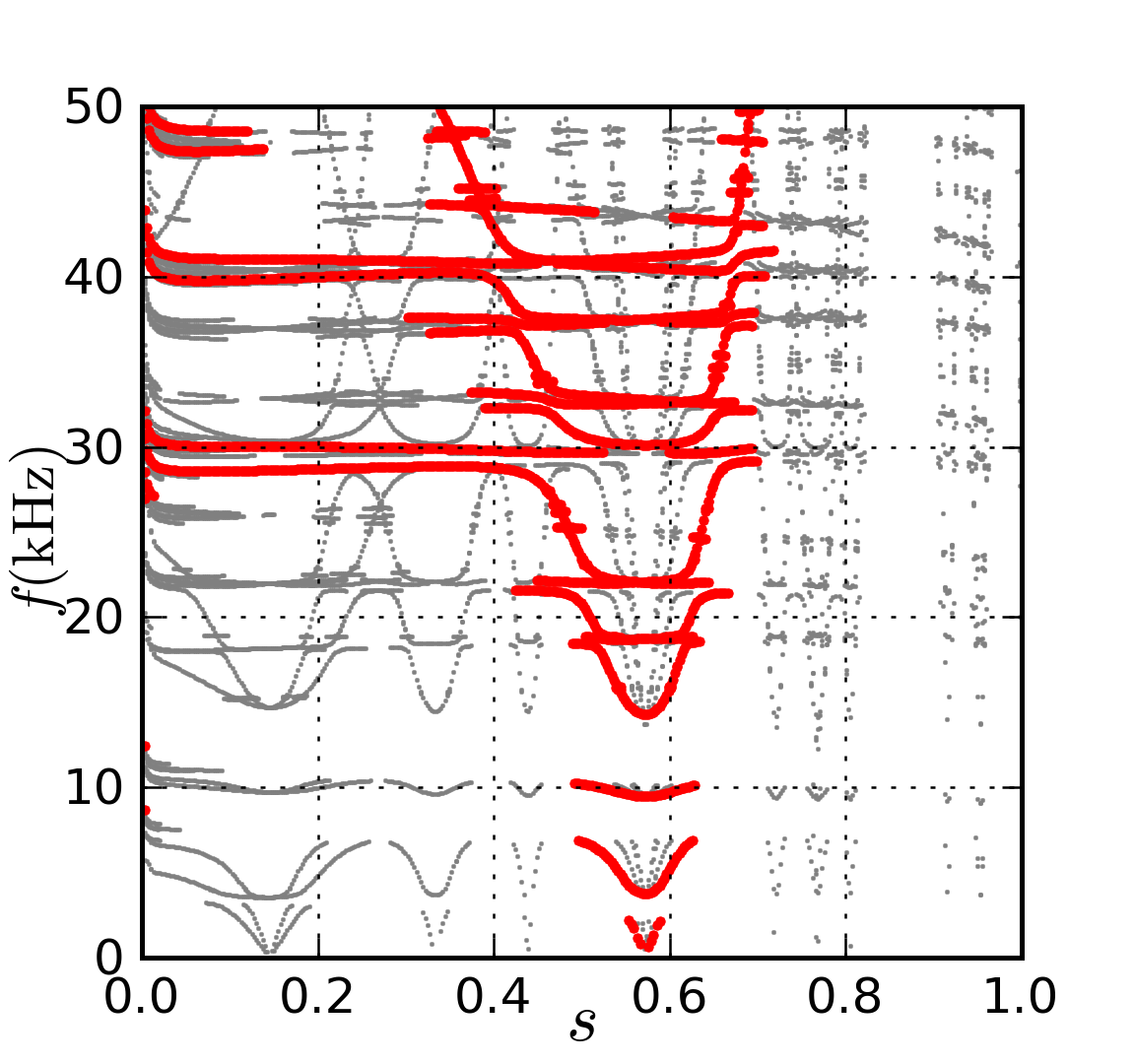}   &
\includegraphics[scale=1]{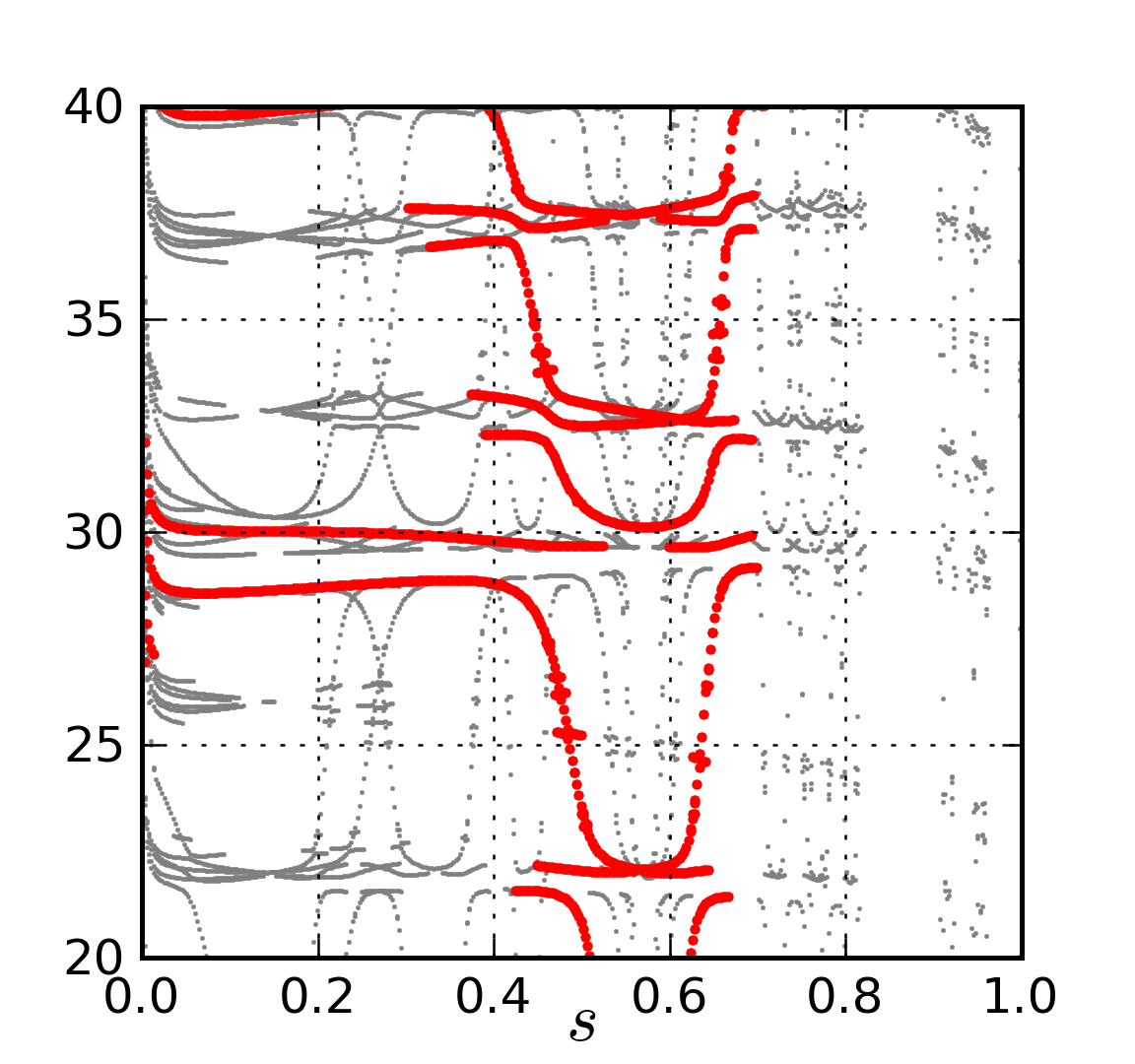}  
\end{array}$$
\caption{\label{fig:spect_low}:  
\quad (a) Low frequency detail of figure \ref{fig:contispect_30_full}, where the $(m,n)=(4,5)$ shear \alf branch is coloured red. (b) Detail of (a).}
\end{figure}

We now turn our attention to the low-frequency domain. Figure \ref{fig:spect_low}a shows the low-frequency detail of figure \ref{fig:contispect_30_full}, where the $(m,n)=(4,5)$ branch has been identified, confirming the significant Alfv\'{e}n-acoustic interactions predicted in subsection \ref{sec:conttheory}. Several Alfv\'{e}n-acoustic gaps are formed, with large $\beta$-induced gaps appearing above some of these notably at $9~\rm{kHz}$, $12~\rm{kHz}$. Another smaller $\beta$-induced gap is found at $30~\rm{kHz}$ which is more easily seen in Figure \ref{fig:spect_low}b. The positive solutions to equation (\ref{eq:gamfreqest}) at $s=0.55$ are $\omega \approx 9.4, 13.2, 31.1~\rm{kHz}$ which are therefore all seen to be approximate geodesic frequency solutions.

\begin{figure}[h]
\centering
$$\setlength\arraycolsep{0 em}
\begin{array}{cc}
(a)&(b)\\
\includegraphics[scale=1]{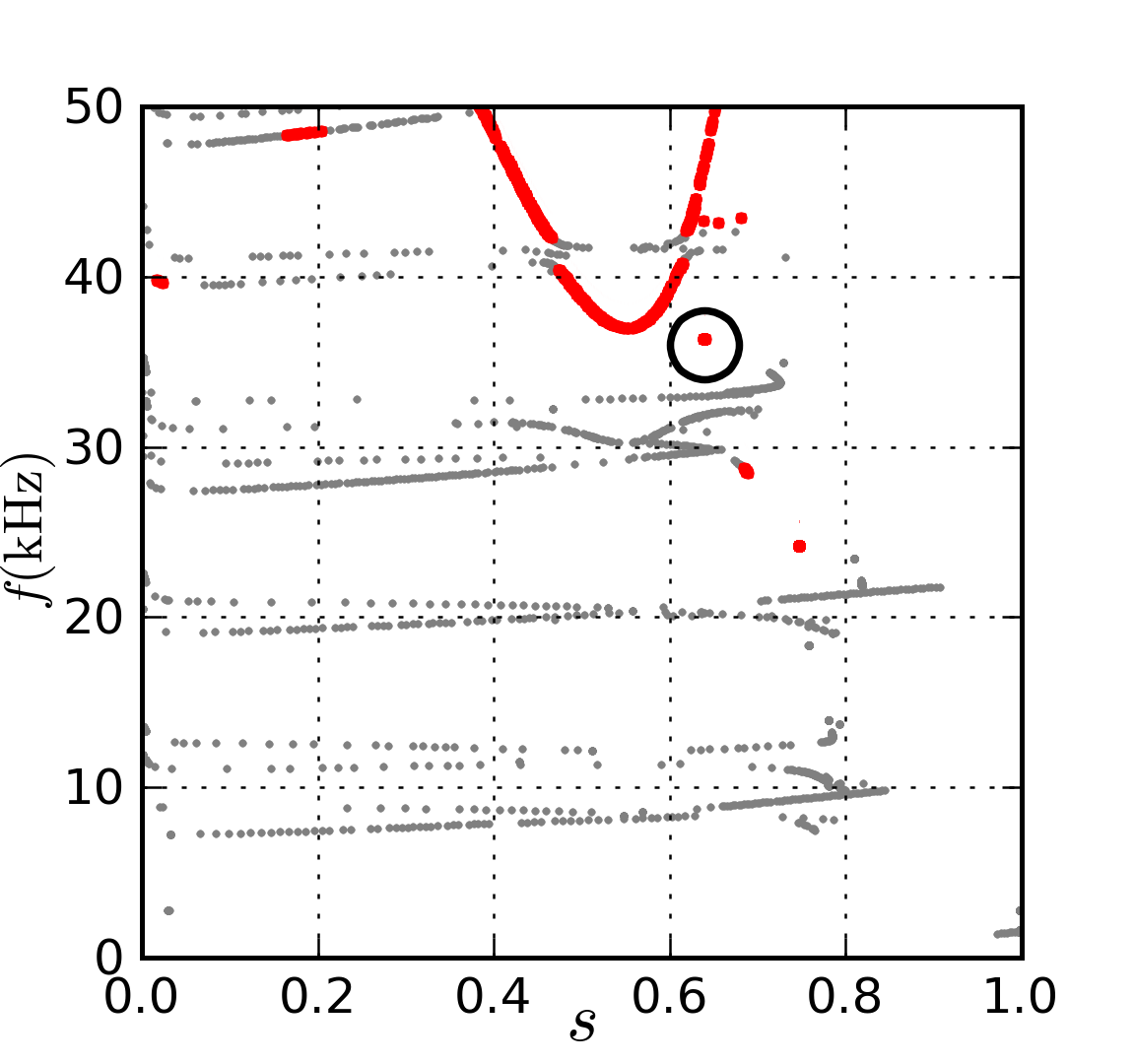}   &
\includegraphics[scale=1]{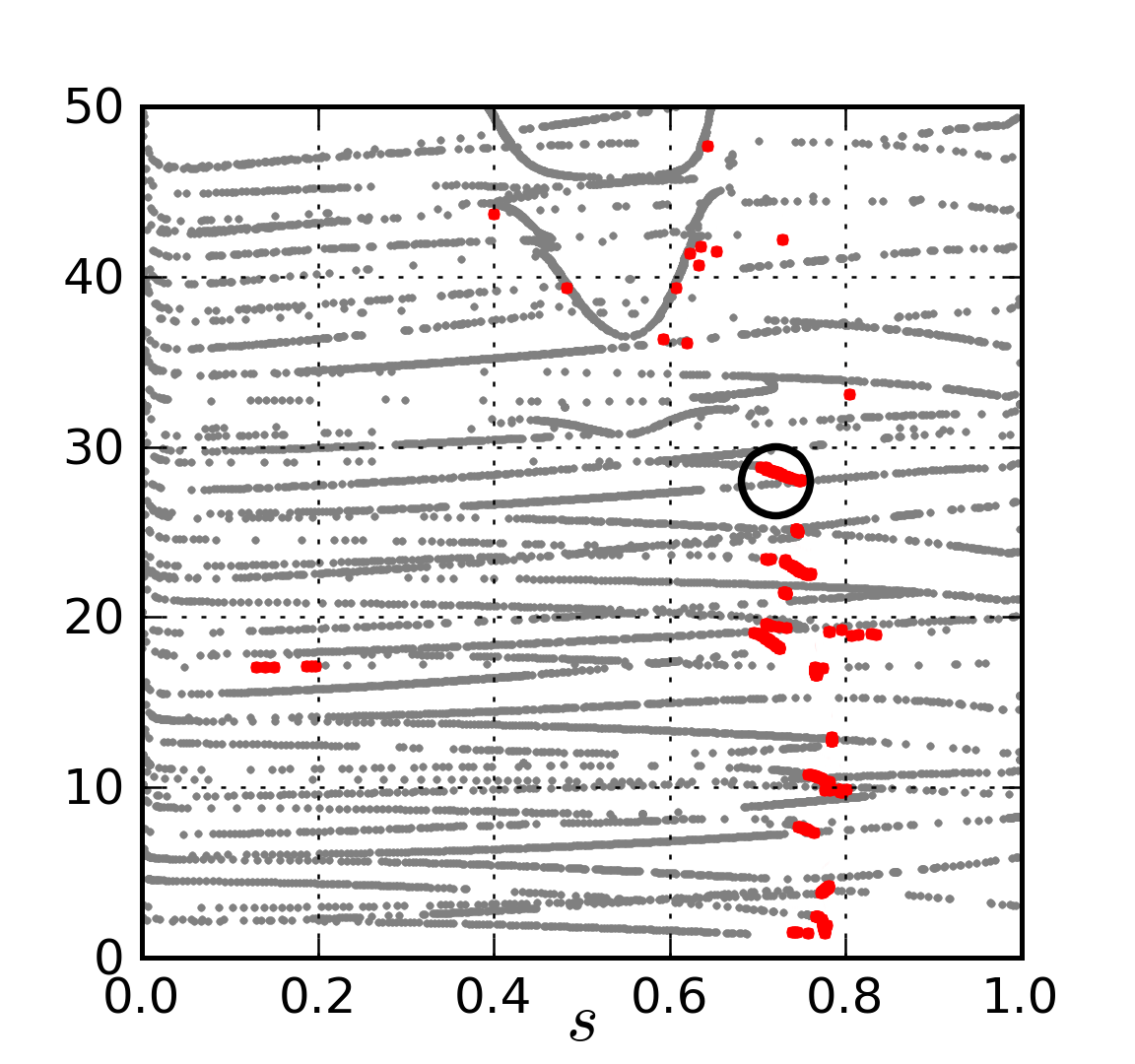} 
\end{array}$$
\caption{\label{fig:cas3dspect_30}  
\quad (a) Low frequency CAS3D spectrum where the $(m,n)=(4,5)$ shear \alf branch is coloured red (polarisation classification). Only adjacent poloidal harmonics and two toroidal sidebands of the $(4,5)$ harmonic included in the Fourier table for all mode vector components (total of $15$ harmonics for each). (b) Same as (a) with larger Fourier table and potential energy classification. The $\bxi_\perp$ components have $16$ near-resonance Fourier harmonics with $m_{\rm{min}}=3$, $m_{\rm{max}}=5$, and $n_{\rm{min}}=-10$ and $n_{\rm{max}}=0$ while the parallel displacement has $m_{\rm{min}}=2$, $m_{\rm{max}}=6$, $n_{\rm{min}}=-14$ and $n_{\rm{max}}=4$. In both figures $1000$ flux surfaces are used, and $\kappa_h=0.30$.}
\end{figure}

Figure \ref{fig:cas3dspect_30}a shows the full $\kappa_h=0.3$ low-frequency ideal MHD spectrum produced by CAS3D, where a small mode-number table (adjacent poloidal mode-numbers and two toroidal sidebands) has been used to make the gap structure clear. The CAS3D spectrum differs from the CONTI spectrum in that it shows a strong geodesic cut-off near $35~\rm{kHz}$ and the lower sections of the shear \alf spectrum are absent. 

\begin{figure}[h]
\centering
$$\setlength\arraycolsep{0 em}
\begin{array}{cc}
(a)&(b)\\
\includegraphics[width=80mm]{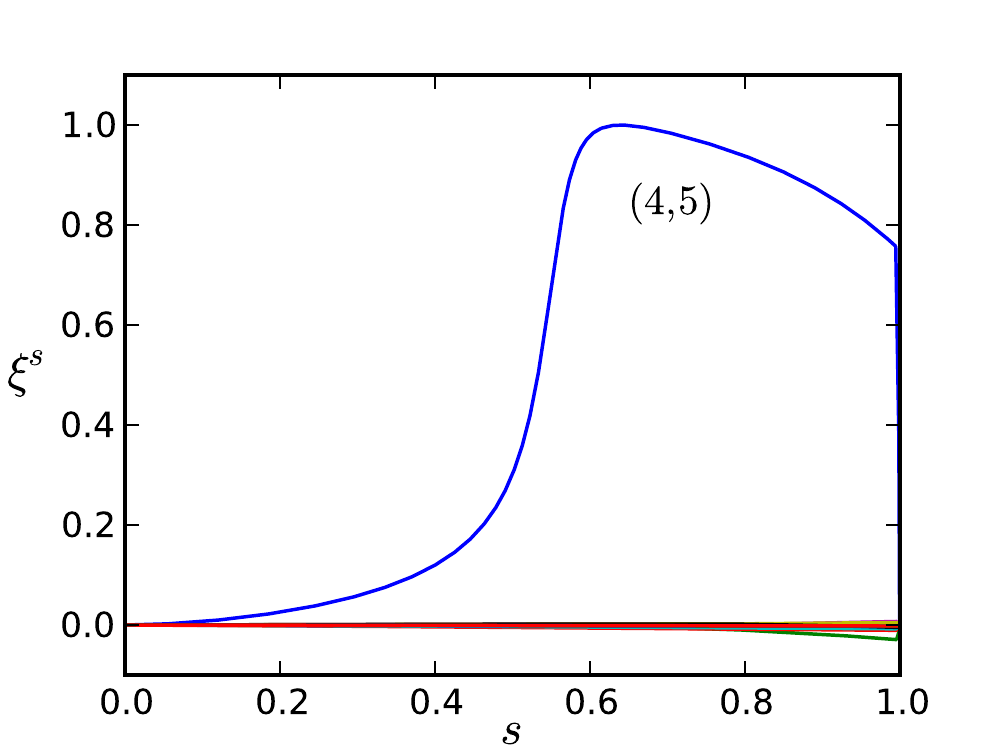}   &
\includegraphics[width=80mm]{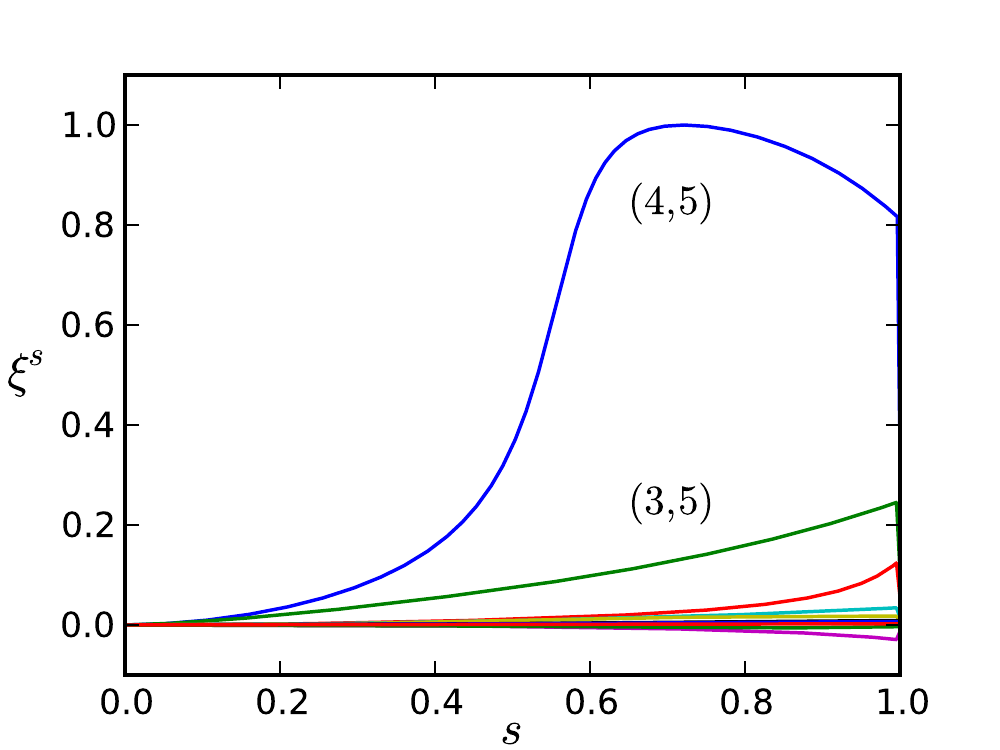} 
\end{array}$$
\caption{\label{fig:gap_modes}  
\quad (a) CAS3D discrete mode circled in Figure \ref{fig:cas3dspect_30}a. (b) CAS3D discrete mode found near cluster circled in Figure \ref{fig:cas3dspect_30}b. Both cases are for $\kappa=0.30$. Some mode harmonics have been labelled and it is seen that both modes are dominated by $(m,n)=(4,5)$. The vertical axis is the normalised fluid displacement orthogonal to the flux surfaces $\xi^s$.}
\end{figure}

Several ``quasi-discrete" modes are found below $35~\rm{kHz}$, meaning that the mode spatial structures have large sections apparently free from continuum interactions. Figure \ref{fig:gap_modes}a shows an example BAE using a small Fourier basis in order to suppress acoustic resonances. Even with the inclusion of many sound branches, eigenmode clusters with shear Alfv\'{e}n characteristics are found as shown in \ref{fig:cas3dspect_30}b using the potential energy classification scheme. Quasi-discrete modes are found in some of these clusters, such as the mode shown in figure \ref{fig:gap_modes}b, found near the cluster circled in \ref{fig:cas3dspect_30}b. As these are fixed-boundary simulations, the fluid displacement orthogonal to the flux surfaces $\xi^s$ must vanish at the edge, which it does due to a sharp gradient discontinuity near the edge presumably due to continuum interactions. Modes with a very similar structure but the discontinuity appearing further inside the plasma are also found. The modes shown in figure \ref{fig:gap_modes} may therefore experience significant continuum damping.

\begin{figure}[h]
\centering
$$\setlength\arraycolsep{0 em}
\begin{array}{cc}
(a)&(b)\\
\includegraphics[width=80mm]{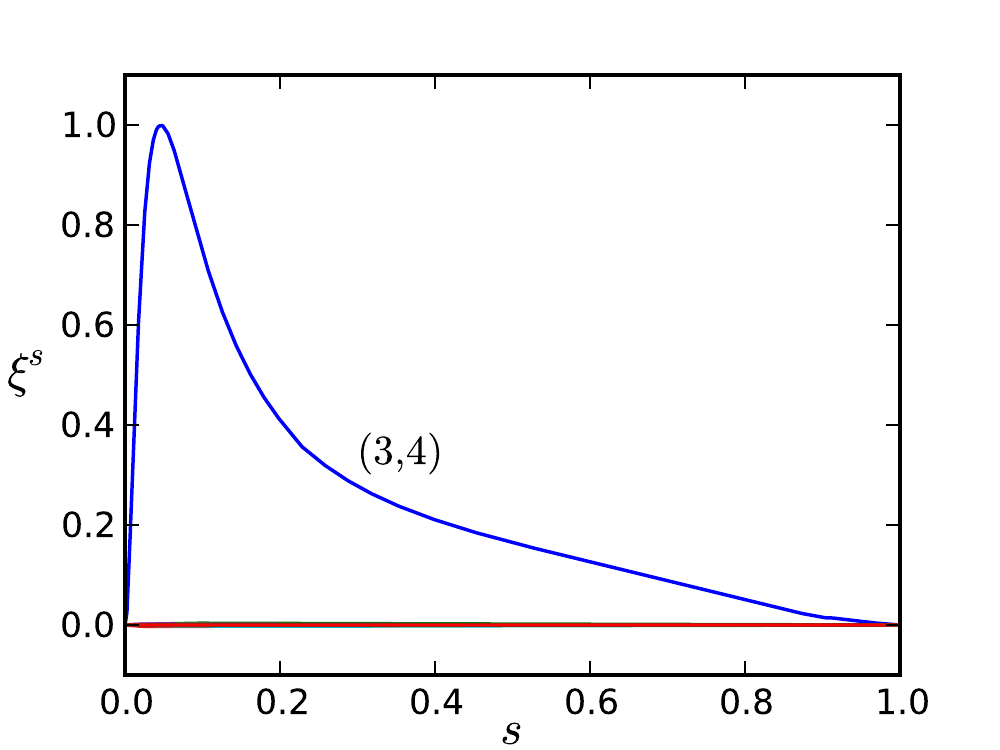}   &
\includegraphics[width=80mm]{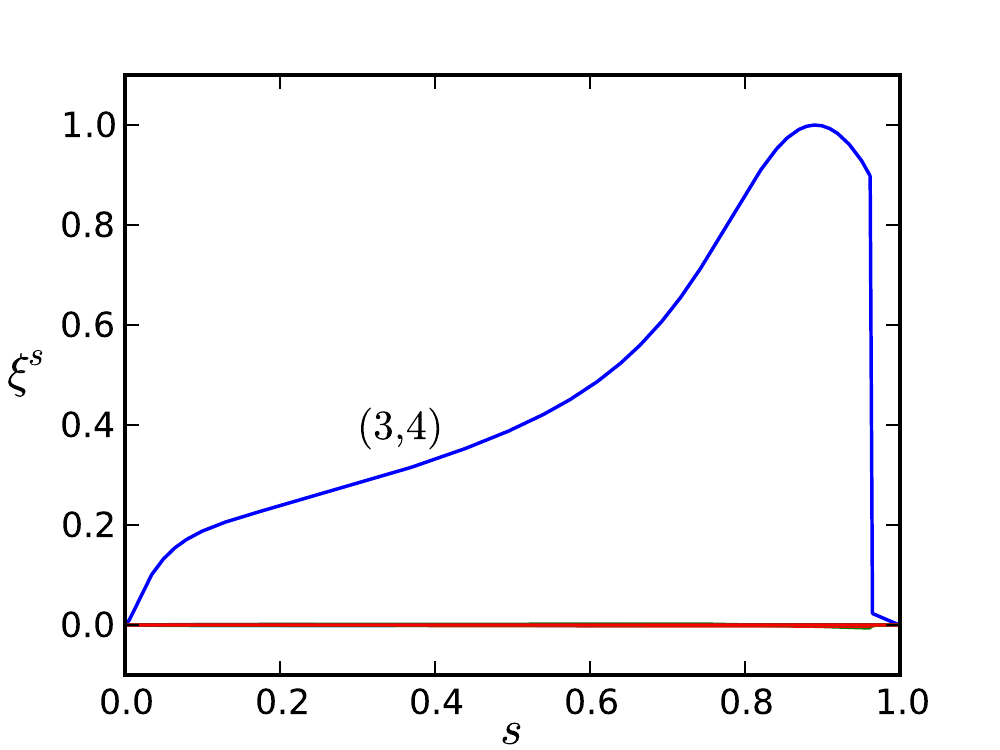} 
\end{array}$$
\caption{\label{fig:gap_modes_60}  
\quad The normalised fluid displacement orthogonal to the flux surfaces $\xi^s$ for: (a) A $(m,n)=(3,4)$ discrete mode at $38~\rm{kHz}$ and (b) a $(m,n)=(3,4)$ quasi-discrete mode at $34~\rm{kHz}$. Both cases are for $\kappa=0.60$.}
\end{figure}

Low-frequency quasi-discrete modes dominated by $(m,n)=(3,4)$ are also found for the left branch of the high-$\kappa_h$ whale-tail. Two $\kappa=0.60$ quasi-discrete modes are shown in figures \ref{fig:gap_modes_60}a and \ref{fig:gap_modes_60}b, with frequencies of $38~\rm{kHz}$ and $34~\rm{kHz}$ respectively. The mode in figure \ref{fig:gap_modes_60}b shows clearly a typical example of the aforementioned near-edge resonances. On the right sides of the whale-tails, near $\kappa=0.50$ and $\kappa=0.90$, we have not been able to find low-frequency Alfv\'{e}nic discrete modes dominated by $(m,n)=(4,5)$ and $(m,n)=(3,4)$ respectively.

\section{Alfv\'{e}n-acoustic whale-tails?}

Since $\beta$-induced gaps below $50~\rm{kHz}$ and quasi-discrete modes with reduced continuum damping exist in these gaps, we offer a partial alternative explanation of H-1 wave activity in terms of $\beta$-induced \alf eigenmodes (BAEs) or $\beta$-induced Alfv\'{e}n-acoustic eigenmodes (BAAEs).

\begin{figure}[h]
\centering
$\begin{array}{cc}
(a)&(b)\\
\includegraphics[width=80mm]{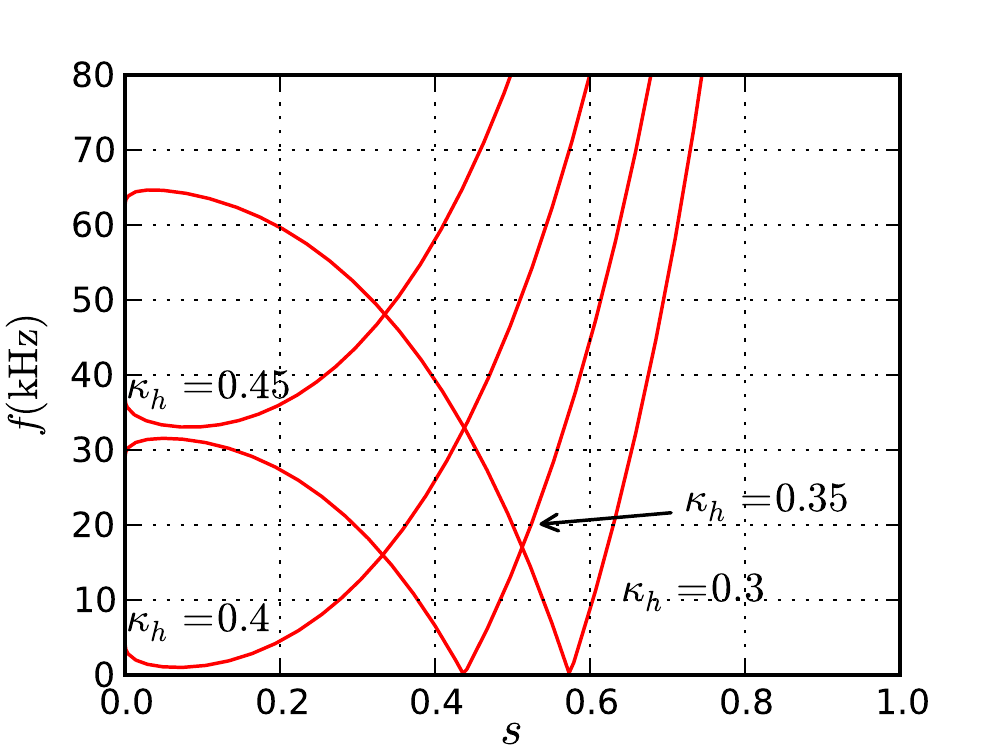} & 
\includegraphics[width=80mm]{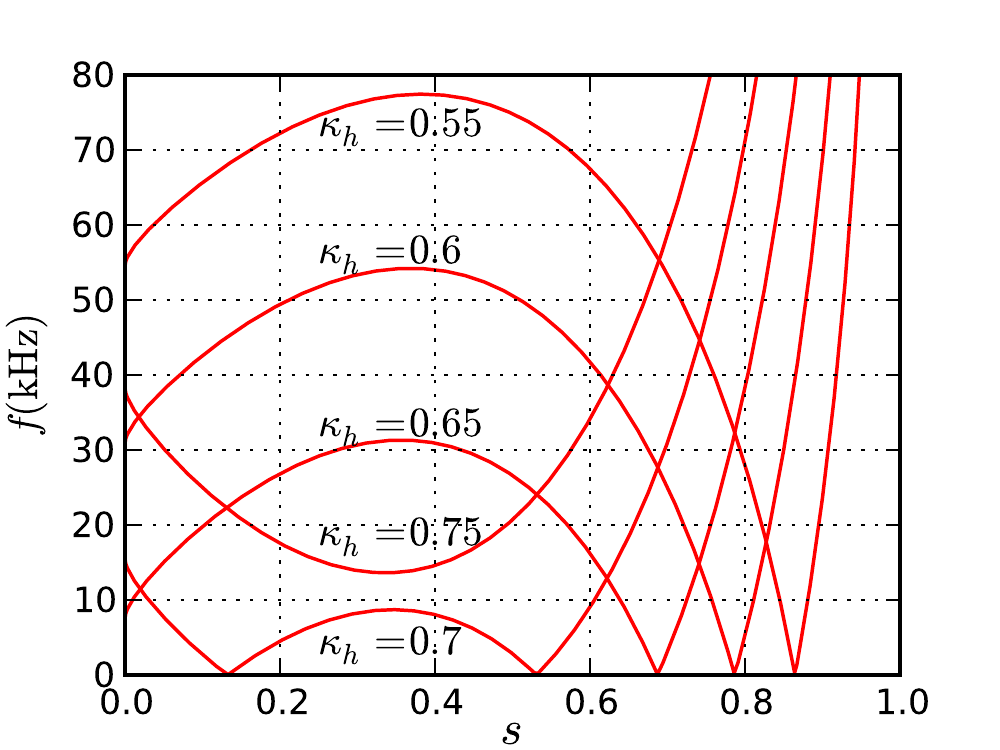}  
\end{array}$
\caption{\label{fig:cylcont} 
(a) Cylindrical shear \alf continua for $(m,n)=(4,5)$ showing $\kappa_h$ dependence near $\kappa_h=0.40$. (b) Cylindrical shear \alf continua for $(m,n)=(3,4)$ showing $\kappa_h$ dependence near $\kappa_h=0.75$.} 
\end{figure}

Figure \ref{fig:cylcont}a shows the $\kappa_h$ dependence of the $(m,n)=(4,5)$ cylindrical shear \alf continuum $\omega_A$ near $\kappa=0.4$ (the low-$\kappa$ whale-tail) and figure \ref{fig:cylcont}b shows the same for $(m,n)=(3,4)$ near $\kappa=0.75$ (the high-$\kappa$ whale-tail). The left branches of both of these whale-tails coincide with the inward migration of the outermost surface at which the cylindrical shear \alf continuum vanishes $s_{\rm{res}}$. The geodesic acoustic minimum of the 3-dimensional compressible shear \alf spectrum is located near $s_{\rm{res}}$ (compare figure \ref{fig:spect_low}a with the $\kappa=0.30$ curve in figure \ref{fig:cylcont}a), and it will migrate inwards with $s_{\rm{res}}$. 

\begin{figure}[h]

\includegraphics[scale=0.8]{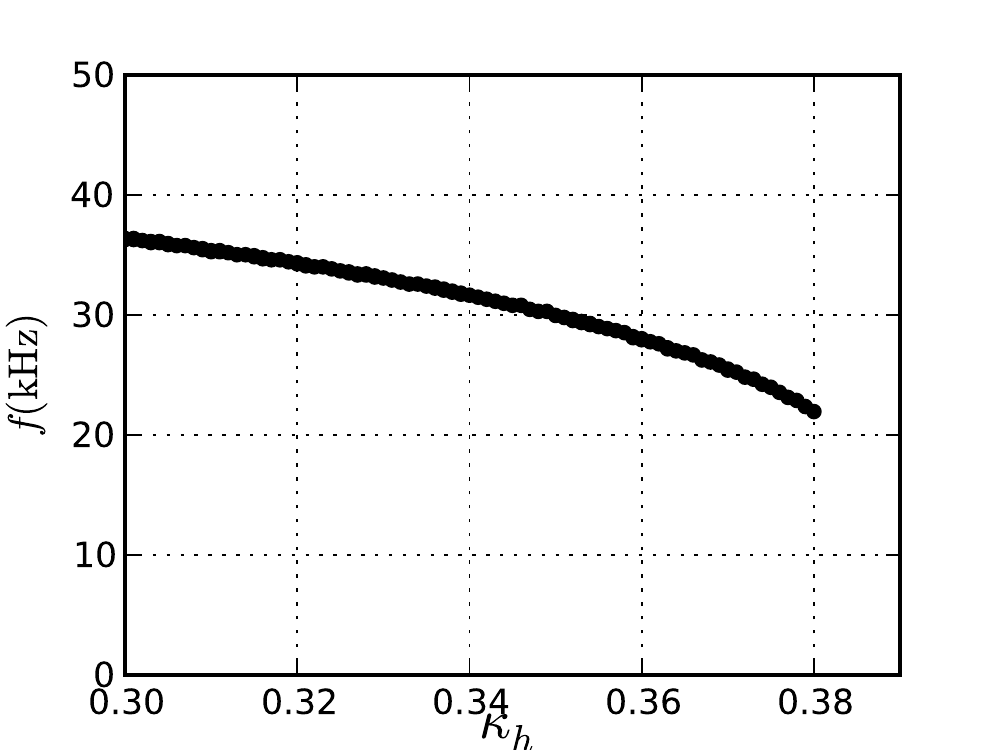}
\centering  
\caption{\label{fig:gamwhale} 
 The $\approx 32~\rm{kHz}$ branch of $\omega_G$ as a function of $\kappa_h$ based on a postulated hollow total temperature profile.} 
\end{figure}

There are indications that H-1 electron temperature profiles increase steeply towards the plasma edge and also that they ``hollow out" as $\kappa_s=0.4$ and $\kappa_s=0.75$ are approached.\cite{oliver_electronically_2008} Consider the left side of the low-$\kappa_h$ whale-tail covering the range $0.3\leq \kappa_h \leq 0.4$. In the absence of ion temperature measurements or accurate spatially-resolved electron temperature profiles for these values of $\kappa_h$, we postulate a total temperature profile $T(\kappa_h,s_{\rm{res}})$ that evolves linearly from $T=27~\rm{eV}$ at $\kappa_h=0.3$ and $s_{\rm{res}}\approx 0.55$ to $T=8~\rm{eV}$ at $\kappa_h=0.38$ and $s_{\rm{res}}\approx 0.25$ where the values $s_{\rm{res}}$ are taken from the $(m,n)=(4,5)$ Alfv\'{e}n branch (compare with figure \ref{fig:cylcont}). We emphasise; this temperature profile is not intended to represent the temperature for a single discharge, it represents local temperatures in $(s,\kappa_h)$ space postulating a simple linear profile consistent with preliminary spatially resolved electron temperature reconstructions\cite{oliver_electronically_2008}, assuming $T_e=T_i$. These temperatures can be used in equation (\ref{eq:gamfreqest}) to estimate how the $\approx 32~\rm{kHz}$ branch of $\omega_G$ evolves with $\kappa_h$ for the $(m,n)=(4,5)$ shear \alf branch in the range $0.3\leq \kappa_h \leq 0.4$, which is shown in figure \ref{fig:gamwhale}. Comparison of figure \ref{fig:gamwhale} with figure \ref{fig:freqvskappa_davethesis} shows that BAEs or BAAEs excited just below $\omega_G$ will show a very similar configurational frequency dependence with the left side of the low-$\kappa$ whale-tail. 

Unfortunately the available electron temperature measurements do not extend to the outer flux surfaces ($s>0.6$) which would be necessary for an explicit comparison of $\omega_G$ in the range $0.6\leq \kappa_h \leq 0.75$ with the left side of the high-$\kappa$ whale-tail. It is clear though that a similar frequency dependence on $\kappa_h$ will be observed provided that the temperature increases accordingly; that it increases to some degree is fairly certain.

The quasi-discrete modes found in section \ref{alf_acoustic} will probably experience enhanced  continuum interactions in the presence of the non-constant temperatures used in this section. Whether these interactions will outweigh the relevant driving mechanisms is an important question that requires more detailed physics than ideal-MHD alone provides.\cite{eremin_beta-induced_2010}

\section{Discussion and Conclusions}

The linearised ideal MHD spectrum is the starting point for analysis of suspected \alf wave observations, as the gap structure of that spectrum identifies frequency ranges with reduced continuum damping. We have presented the three-dimensional, compressible ideal spectrum for H-1 plasmas. The presence of $\beta$-induced gaps and partial gaps, the existence of discrete modes in these gaps and the predicted configurational dependence of the highest BAE gap frequency in the presence of H-1's hollow temperature profiles can account for important features of wave frequency measurements, namely the left sides of the V-shaped ``whale-tail" frequency structures observed in configuration space. Thus, a potentially complete explanation of the left sides of both whale-tails has been provided.

This ``BAE hypothesis" does not, however, provide an explanation for the right sides of the whale-tails as the relevant resonant surface will leave the plasma and the \alf continuum will rapidly approach higher frequencies. For example, at $\kappa_h=0.54$ the $(4,5)$ minimum frequency (see Fig. \ref{fig:contispect_30_full}b) is more than twice as large as measurements. Furthermore, no discrete modes were found below the $(4,5)$ and $(3,4)$ shear \alf continuum minima near $\kappa=0.45$ and $\kappa=0.9$ (the right-hand sides of the whale-tails). Nevertheless, as quasi-discrete Alfv\'{e}nic modes with dominant mode numbers $(4,5)$ and $(3,4)$ could not be found at low-frequencies for $\kappa=0.45$ and $\kappa=0.9$ respectively, there do not appear to be any ideal MHD candidates for the right sides of the whale-tails apart from GAEs that have had their frequencies lowered by non-ideal effects, or by an effective mass $m_{\rm{eff}}$ around six times greater near the plasma core than our impurity-free assumption of $m_{\rm{eff}}=2.5 m_{\rm{p}}$ where $m_{\rm{p}}$ is the proton mass. The factor of six follows from the GAE-measurement discrepancy factor which we showed was reduced to $\lambda\approx 2.5$ when 3-dimensional effects are included.

GAE and BAE wave physics should be experimentally distinguishable by the fact that only BAEs are strongly affected by plasma temperature. This highlights the importance of reliable spatially-resolved temperature measurements which are the subject of future work. It is important that uncertainty in the effective mass and its radial dependence be reduced, as this parameter affects both shear \alf and acoustic frequency predictions. It is hoped that measurements from the new helical Mirnov array have the potential to provide multiple mode-number assignments as well as eventually allowing for comparison with eigenmode polarisation predictions. It may also be possible to obtain accurate spatially resolved measurements of the relative plasma density fluctuations which, when compared with Mirnov signals would help in characterising shear \alf versus acoustic components of the observed wave dynamics. Ultimately, candidate modes should be compared with Mirnov array signals and internal measurements for a conclusive identification.

Efforts are also underway to obtain energetic particle drive estimates for CAS3D eigenfunctions and extend equilibrium and spectral calculations to incorporate a free boundary and vacuum field. Ideal MHD must be interpreted with due caution at low frequencies due to the poor description it provides of compressible motions\cite{freidberg_ideal_1982}, and the omission of thermal ion interactions, such as ion Landau damping. For these reasons it may prove fruitful to pursue kinetic models of sub-GAM H-1 wave physics.\cite{kolesnichenko_sub-gam_2008,gorelenkov_beta-induced_2009} 

\section*{Acknowledgements}

We wish to thank Carolin N\"{u}hrenberg for providing CAS3D and for many insightful discussions, and Axel Konies for making CONTI available to us and for informative email correspondence. We also thank Shaun Haskey for providing well-organised magnetic fluctuation data. The authors would like to acknowledge support from the  Australian Research Council through grants FT19899 (MJH) and DP0451960 (BDB), and the Australian Government's Major National Research Facility Scheme.

\bibliographystyle{unsrt}
{\footnotesize
\bibliography{ref}}

\end{document}